\documentclass[twocolumn,prl,showpacs,amsmath,amssymb]{revtex4}

\usepackage{graphicx}

\def\be{\begin{equation}}
\def\ee{\end{equation}}
\def\bea{\begin{eqnarray}}
\def\eea{\end{eqnarray}}
\def\bma{\begin{mathletters}}
\def\ema{\end{mathletters}}

\def\0{\overline{0}}

\def\q0{\underline{0}}

\def\tr{\mbox{tr}}
\def\one{\leavevmode\hbox{\small1\normalsize\kern-.33em1}}
\def\compl{\begin{picture}(8,8)\put(0,0){C}\put(3,0.3){\line(0,1){7}}\end{picture}}

\def\bra#1{\langle#1|} \def\ket#1{|#1\rangle}
\def\braket#1#2{\langle#1|#2\rangle}

\def\proj#1{\ket{#1}\!\bra{#1}}

\begin{document}

\title{
Equivalence between two-qubit entanglement and secure key
distribution }

\author{Antonio Ac\'\i n$^{1,2}$, Lluis Masanes$^3$
and Nicolas Gisin$^1$}

\affiliation{ $^1$GAP-Optique, University of Geneva, 20, Rue de
l'\'Ecole de
M\'edecine, CH-1211 Geneva 4, Switzerland\\
$^2$Institut de Ci\`encies Fot\`oniques, Jordi Girona 29, Edifici Nexus II, 08034 Barcelona, Spain\\
$^3$Dept. ECM, University of Barcelona, Diagonal 647, 08028
Barcelona, Spain}

\date{\today}


\begin{abstract}

We study the problem of secret key distillation from bipartite
states in the scenario where Alice and Bob can only perform
measurements at the single-copy level and classically process the
obtained outcomes. 
Even with these limitations, secret bits can be asymptotically
distilled by the honest parties from any two-qubit entangled
state, under any individual attack. Our results point out a
complete equivalence between two-qubit entanglement and secure key
distribution: a key can be established through a one-qubit channel
if and only if it allows to distribute entanglement. These results
can be generalized to higher dimension for all those states that
are one-copy distillable.

\end{abstract}


\maketitle

Quantum correlations or entanglement is the basic ingredient for
many applications of Quantum Information Theory \cite{book}. By
exploiting the correlations of entangled states,  
one can perform tasks that are impossible in Classical Information
Theory. Quantum cryptography \cite{review}, or more precisely
quantum key distribution, is the most successful Quantum
Information application, due to its experimental feasibility with
present-day technology. Although entanglement is not required for
a secure key distribution \cite{BBM}, there exist proposals using
entangled states \cite{Ekert}. Indeed, it is unclear which role
entanglement plays in quantum cryptography protocols. In this
work, we analyze the problem of secret key extraction in the
following scenario: after a distribution stage, two honest
parties, Alice and Bob, share a quantum state. This state is
translated into a probability distribution by local measurements
at the single-copy level, and the obtained outcomes are processed
in order to distill a secret key. We denote by SIMCAP this {\sl
SIngle-copy Measurements plus ClAssical Processing} scenario. This
is a common scenario in Quantum Information applications, where
useful correlations are distributed between two or more parties by
means of entangled states. For two-qubit systems and individual
attacks, we prove that Alice and Bob can distill a key by a SIMCAP
protocol if and only if they initially share an entangled state.
Thus, two-qubit entanglement is indeed equivalent to secure key
distribution.

Our result links the security of one-qubit channels with their
entanglement capability. In the usual formulation of Quantum
Cryptography, first a protocol for key distribution is proposed
and later possible eavesdropping attacks on it are analyzed.
However, one can reverse this standard presentation and, after
specifying an eavesdropping attack, look for a secure key
distribution protocol. This is indeed closer to what happens in a
practical situation: the honest parties are connected by a given
channel, denoted by $\Upsilon$, that is fixed and known. It
depends on experimental parameters such as, for instance, dark
counts or optical imperfections, and is the only non-local quantum
resource Alice and Bob share. From the quantum cryptography point
of view, it is conservatively assumed that Eve has total access to
the channel. This means that the definition of the quantum channel
is equivalent to specify Eve's interaction with the sent states.
When does a given channel allow the honest parties to securely
establish a secret key, in the SIMCAP scenario? Our results imply
that a one-qubit channel is secure as soon as it allows
entanglement distribution. For any entangling channel we show how
to construct the corresponding SIMCAP key distillation protocol.
Moving to higher dimension, our results immediately hold for all
those bipartite states, and corresponding channels, that are
one-copy distillable. Thus, they suggest a complete equivalence
between distillable entanglement and secure key distribution.


Let us start with the simplest case of two qubits. A two-qubit
entangled state
is locally prepared by Alice and one of the two qubits is sent to
Bob through a quantum channel. Since the channel is not perfect,
Alice and Bob end with a two-qubit mixed state, $\rho_{AB}$
\cite{notesource}. They attribute the channel imperfections to the
eavesdropper, Eve, who interacts with the sent qubits. We assume,
as it is often done in many works on Quantum Cryptography, that
Eve applies an individual attack: she lets independent auxiliary
systems interact with each qubit and measures each system before
the key extraction process \cite{noteind}. Since Eve has a perfect
control on her interaction, the global state of the system is
pure, $\ket{\Psi_{ABE}}$. The state shared by Alice and Bob is the
one resulting from tracing out Eve,
$\rho_{AB}=\tr_E(\ket{\Psi_{ABE}}\!\bra{\Psi_{ABE}})$. The global
pure state including Eve is, without loss of generality,
\begin{equation}\label{psiabe}
    \ket{\Psi_{ABE}}=\sum_{i=1}^r \sqrt{r_i}\,\ket{i}\ket{i_e} ,
\end{equation}
where $\ket{\Psi_{ABE}}\in\compl^2\otimes\compl^2\otimes\compl^r$,
$\{r_i,\ket{i}\}$ define the spectrum of $\rho_{AB}$, $r$ is
its rank and $i_e$ is an orthonormal basis on Eve's space. 
By computing the Schmidt decomposition 
with respect to the partition $AB-E$, one can easily see that any
other state
$\ket{\tilde\Psi_{ABE}}\in\compl^2\otimes\compl^2\otimes\compl^{d_E}$,
where $d_E\geq r$, such that
$\tr_E(\ket{\tilde\Psi_{ABE}}\!\bra{\tilde\Psi_{ABE}})=\rho_{AB}$,
is completely equivalent to $\ket{\Psi_{ABE}}$.

If $\rho_{AB}$ is entangled, one can consider the following {\sl
fully quantum} protocol for key distribution. The honest parties
run a quantum distillation protocol \cite{dist} that transforms
many copies of the initial mixed entangled state into fewer copies
of a maximally entangled state \cite{horo}. In this way, Eve
becomes uncorrelated to Alice and Bob, who can safely measure in
one basis, say $z$, and obtain the secret key. Note that in these
protocols the honest parties must be able to perform quantum
operations on several copies of their local states. This is in
strong contrast to the SIMCAP scenario where all the collective
actions are performed at the classical level, while quantum
physics is only used for the correlation distribution. Does this
limit the possibility of distilling a key?

It is worth to mention here that the experimental requirements for
the SIMCAP protocols are definitely less stringent than for
quantum distillation protocols. In particular, no quantum memory
is needed, avoiding decoherence problems. Moreover, our scenario
reflects precisely what is feasible with current technology, in
contrast to joint operations and quantum memories, that are
impossible on a large scale even in the near future.



{\bf Theorem:} {\sl Consider the situation in which Alice and Bob
share unlimited many instances of a two-qubit state, $\rho_{AB}$.
Under individual attacks, they can distill a secret key from them
by measurements at the single-copy level and classical processing
of the outcomes if and only if $\rho_{AB}$ is entangled.}


{\em Proof:} 
It was shown in \cite{optfilt} that there exists a unique local
filtering operation, $F_A\otimes F_B$ with $F_A^\dagger
F_A\leq\one_2$ and $F_B^\dagger F_B\leq\one_2$, mapping with some
probability any two-qubit state into a state diagonal in a Bell
basis \cite{bellbasis},
\begin{eqnarray}\label{rhod}
    \rho'_{AB}&=&\Lambda_1\ket{\Phi^+}\!\bra{\Phi^+}
    +\Lambda_2\ket{\Psi^+}\!\bra{\Psi^+}\nonumber\\
    &+&\Lambda_3\ket{\Psi^-}\!\bra{\Psi^-}
    +\Lambda_4\ket{\Phi^-}\!\bra{\Phi^-} .
\end{eqnarray}
The local bases can be chosen such that $\ket{\Phi^+}$ is the
eigenvector associated to the largest eigenvalue, 
$\Lambda_1=\max(\{\Lambda_i\})$. 
This transformation maps entangled states into entangled Bell
diagonal states \cite{optfilt}. After applying this filtering
operation to $\rho_{AB}$, the honest parties share a Bell diagonal
state (\ref{rhod}), while the global state is
\begin{eqnarray}\label{psibell}
    \ket{\Psi'_{ABE}}&=&\lambda_1\ket{\Phi^+}\ket{1}
    +\lambda_2\ket{\Psi^+}\ket{2}\nonumber\\
    &+&\lambda_3\ket{\Psi^-}\ket{3}
    +\lambda_4\ket{\Phi^-}\ket{4} ,
\end{eqnarray}
with $\lambda_i=\sqrt{\Lambda_i}$. Since the positivity of the
partial transposition \cite{parttr} is a necessary and sufficient
condition for separability in $\compl^2\otimes\compl^2$ systems,
$\rho'_{AB}$ is entangled {\sl iff}
\begin{equation}\label{rhoent}
    \Lambda_1>\Lambda_2+\Lambda_3+\Lambda_4=1-\Lambda_1 .
\end{equation}

After a successful local filtering, Alice and Bob measure
$\rho'_{AB}$ in the $z$ basis, obtaining a partially correlated
list of symbols, $\{a_i\}$ and $\{b_i\}$. The measurements in the
$z$ basis terminate the measurement step in the SIMCAP
distillation protocol, after which the original quantum state has
been translated into a probability distribution \cite{notemeas}.
From Eq. (\ref{psibell}), Eve's non-normalized states,
$\ket{e_{AB}}$, depending on Alice and Bob's results are, where
$R=0,1$,
\begin{eqnarray}\label{evest}
  &&\ket{e_{RR}}=\frac{1}{\sqrt{2}} (\lambda_1\ket{1} +(-1)^R \lambda_4\ket{4})
  \nonumber\\
  &&\ket{e_{R(1-R)}}=\frac{1}{\sqrt{2}} (\lambda_2\ket{2} +(-1)^R
  \lambda_3\ket{3}).
\end{eqnarray}
Note that Eve knows in a deterministic way whether Alice and Bob
differ in their measurement outcomes (which implies
$I_{AE}=I_{BE}$). This happens with probability
\begin{equation}\label{errab}
    \epsilon_{B}=\|e_{01}\|^2+\|e_{10}\|^2=\Lambda_2+\Lambda_3 ,
\end{equation}
which is Bob's error probability.

In order to classically distill a key, Alice and Bob will now
apply the advantage distillation protocol described in Ref.
\cite{GW} to their measurement outcomes. If the state is close to
 $\ket{\Phi^+}$, the mutual information between the honest
parties, $I_{AB}$, is larger than Eve's information,
$I_E=\min(I_{AE},I_{BE})$. Then, no advantage distillation
protocol is in principal required, since privacy amplification
\cite{CK}, a more efficient key distillation protocol, suffices.
Nevertheless, we deal with advantage distillation protocols
because they allow to extract a key even in situations where
$I_{AB}\leq I_{E}$ \cite{Maurer}. The advantage distillation
protocol works as follows: if Alice wants to establish the bit $x$
with Bob, she randomly takes $N$ items from her list of symbols,
$\vec a=(a_1,a_2,\ldots,a_N)$, and sends to Bob the vector $\vec
x$ such that $a_i+x_i=x \mbox{ mod }2,\,\forall\, i=1,\ldots,N$,
plus the information about the chosen symbols. Bob computes
$b_i+x_i$, and whenever he obtains the same result,
$b_i+x_i=y,\,\forall\, i$, he accepts the bit. If not, the symbols
are discarded and the process is repeated for a
new vector of length $N$. 
Bob's error probability is now \cite{GW}
\begin{equation}\label{errABN}
    \epsilon_{BN}=\frac{(\epsilon_{B})^N}{(1-\epsilon_{B})^N+
    (\epsilon_{B})^N}\leq\left(\frac{\epsilon_{B}}
    {1-\epsilon_{B}}\right)^N ,
\end{equation}
that tends to an equality for $N\rightarrow\infty$.

\begin{figure}
  \includegraphics[width=6.5 cm]{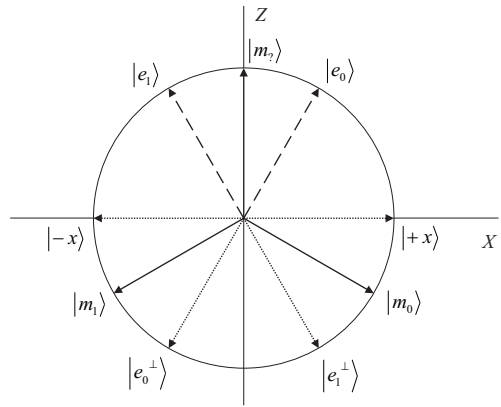}\\
  \caption{Example of measurement attaining
  the bound of Eq. (\ref{expterm}), where the
  first outcome, $\ket{m_0}$, is associated to 0, the second,
  $\ket{m_1}$, symmetric to $\ket{m_0}$ with respect to the
  $z$ axis, to 1 and the third, $\ket{m_?}=\ket{+z}$, to an
  inconclusive result. The weight of $\ket{m_?}$ is minimized.
  One can consider similar three-outcome measurements,
  just changing the angle between $\ket{m_0}$, or $\ket{m_1}$,
  and the $z$ axis. All the measurements such that $\ket{m_0}$
  is between $\ket{+x}$ and $\ket{e^\perp_1}$
  saturate the exponential bound. This does not mean that
  $\epsilon_{EN}$ is the same for all of them. The limiting cases,
  $\ket{m_0}=\ket{e^\perp_1}$ and $\ket{m_0}=\ket{+x}$, correspond
  to the optimal measurements for unambiguous discrimination and
  for maximizing Eve's information.
  }\label{figmeas}
\end{figure}

Notice that for large $N$, $x=y$ with very high probability. Hence
we concentrate on the states $\ket{e_{i}}\equiv
\ket{e_{ii}}/\|e_{ii}\|$, and denote $E_i$ the corresponding
projectors.
Eve applies a generalized measurements (POVM) of $M$ outcomes,
$\sum_i M_{i}=\one_2$ with $M_i>0$, trying to acquire information
about these states.
Indeed, since $\vec a$ is chosen at random, we assume Eve's
measurement to be the same for all qubits without loosing
generality. Moreover, any generic measurement can be seen as a
measurement consisting of rank-one operators where some of the
outcomes are later combined, so we can take
$M_i=\ket{m_i}\!\bra{m_i},\forall\,i$, with
$\|m_i\|\leq 1$. 
After the measurements, Eve uses all the information collected
from the $N$ symbols for guessing $x$. From $\vec x$, she knows
that the bit string was equal to $\vec a=(a_1,a_2,\ldots,a_N)$ or
to $\vec a'=(1-a_1,1-a_2,\ldots,1-a_N)$, corresponding to $1-x$.
Independently of her decision strategy, there are instances where
she will make an error. For example, when the number of zeros in
$\vec a$ is the same as the number of ones (the same holds for
$\vec a'$), and the number of times any measurement outcome has
been obtained is the same for zeros and ones \cite{note2}. These
events do not give her any information about $x$, so she is forced
to guess and makes a mistake with probability 1/2. Therefore, her
error probability is bounded by
\begin{widetext}
\begin{equation}\label{eveerror}
    \epsilon_{EN}\geq\frac{1}{2}\,\frac{1}{2^N}\sum_{n_1,\ldots,n_{M}}
    \frac{N!}{(2n_1)!\ldots (2n_M)!}
    \begin{pmatrix}2n_1 \cr n_1\end{pmatrix}
    \tr(E_0M_1)^{n_1}\tr(E_1M_1)^{n_1}\cdots
    \begin{pmatrix}2n_M \cr n_M\end{pmatrix}
    \tr(E_0M_M)^{n_M}\tr(E_1M_{M})^{n_M} ,
\end{equation}
\end{widetext}
with $2\sum_i n_i=N$. The factor $1/2^N$ takes into account the
number of possible vectors $\vec a$, while the combinatorial terms
count the number of vectors satisfying our requirements. When $N$
is large, one can approximate the combinatorial term
$(2n_i)!\,/(n_i!\,)^2\simeq 2^{2n_i}$ and then
\begin{equation}\label{eveerror2}
    \epsilon_{EN}\gtrsim\frac{1}{2}
    \sum_{n_i}\frac{N!}{(2n_1)!\ldots (2n_M)!}\,
    \prod_{i=1}^M\left(\tr(E_0M_i)\tr(E_1M_i)\right)^{n_i} .
\end{equation}
In the same limit, this sum is equal to
\begin{equation}\label{eveerror3}
    \epsilon_{EN}\gtrsim\frac{1}{2}\,
    \frac{1}{2^{M-1}}
    \left(\sum_{i=1}^M\sqrt{\tr(E_0M_i)\tr(E_1M_i)}\right)^N .
\end{equation}
Since $M_i$ are rank-one operators,
\begin{equation}\label{expterm}
    \sum_{i=1}^M \sqrt{\tr(E_0M_i)\tr(E_1M_i)}=
    \sum_{i=1}^M |\bra{e_0}M_i\ket{e_1}|\geq|\braket{e_0}{e_1}| ,
\end{equation}
where in the last step we used that $\{M_i\}$ is a resolution of
the identity. These equations imply that, for large $N$, Eve's
error probability is bounded by an exponential term
$|\braket{e_0}{e_1}|^N$. This bound is tight: a simple measurement
in the $x$ (i.e. $(\ket{1}\pm\ket{4})/\sqrt 2$) basis attains it
(see Fig. \ref{figmeas} and the appendix).

Now, Alice and Bob can establish a key whenever
\begin{equation}\label{advdist}
    \frac{\epsilon_{B}}{1-\epsilon_{B}}<
    |\langle e_{1}\ket{e_{0}}|
\end{equation}
since then (see Eq. (\ref{errABN})) Bob's error probability
decreases exponentially faster than Eve's, and this condition is
known to be sufficient for key distillation \cite{MW}.
More precisely: if Eq. (\ref{advdist}) is satisfied, there exists
a finite $N$ such that Alice and Bob, starting from the raw data
and using this protocol, end with a smaller list of symbols where
$I_{AB}>I_E$. Then, they can apply privacy amplification
techniques \cite{CK} and distill a key. Using Eqs. (\ref{evest})
and (\ref{errab}), condition (\ref{advdist}) can be shown to be
equivalent to Eq. (\ref{rhoent}). Since Alice and Bob cannot
establish a key when the state $\rho_{AB}$ is separable
\cite{GW2}, we conclude that a secret key can be distilled in the
SIMCAP scenario if and only if the initially shared state is
entangled. $\Box$

Our results imply the equivalence between entanglement and
security for qubit channels: if a one-qubit channel, $\Upsilon$,
allows to distribute entanglement, key distribution is possible.
Indeed, this means that there exists a bipartite state,
$\ket{\Phi}\in\compl^2\otimes\compl^2$, such that 
\begin{equation}\label{entch}
    \rho^{\Phi}_{AB}=(\one_2\otimes\Upsilon)(\ket{\Phi})
\end{equation}
is entangled. Alice can then prepare the state $\ket{\Phi}$
locally and send half of it to Bob through the noisy channel
$\Upsilon$. After this distribution stage, the honest parties
run the presented SIMCAP protocol and distill a secret key from
$\rho^{\Phi}_{AB}$. Two points deserve to be mentioned here.
First, note that if one places the state preparation on Alice's
side,
she can start with the
state ``as if it had passed her filter", i.e. $F_A=\one_2$. 
And second, there is actually no need of entanglement in the
protocol. Indeed, it can be translated into an equivalent protocol
without entanglement using the same ideas as in Ref. \cite{BBM}.
Alice's measurement can be incorporated into the state
preparation, 
before the 
state distribution \cite{note}. Then, she sends through the
channel, with probability $1/2$, one of the two states
$\ket{\psi_B^\pm}\in\compl^2$, defined as
\begin{equation}\label{states}
    \ket{\psi^\pm_B}=\sqrt 2\,(\bra{\pm z}\otimes\one_2)\,\ket{\Phi} .
\end{equation}
Bob receives the states $\rho^\pm_B=\Upsilon(\ket{\psi^\pm_B})$.
He applies the filter $F_B$ and measures in the $z$ basis. Of
course, the obtained probabilities are exactly the same as in the
SIMCAP protocol using $\ket{\Phi}$, so Alice and Bob can securely
distill a key without using any entanglement.

For all the protocols, with and without entanglement, it is
assumed that the channel is fixed. Note that for some channels,
the states $\ket{\psi^\pm_B}$ may be orthogonal and form a basis.
Eve could then replace her interaction by an intercept-resend
attack: she measures in that basis and prepares a new state for
Bob. But this would dramatically change the channel. Thus, Alice
and Bob should randomly interrupt the key distribution and switch
to a check stage where they monitor the channel. Entanglement
is not required for this stage either. 
Those channels that do not allow to distribute entanglement are
called {\sl entanglement breaking}. 
They can be written as \cite{HSR}
\begin{equation}\label{entbr}
    \Upsilon(\ket{\psi})=\sum_k \tr(L_k\ket{\psi}\!\bra{\psi})\,\rho_k ,
\end{equation}
where $\rho_k$ are density matrices and $\{L_k\}$ define a
generalized measurement, i.e. $L_k\geq 0$ and $\sum_k L_k=\one_2$.
From a cryptography point of view, this just represents an
intercept-resend attack, as the one described above.

To conclude, we have seen that, under arbitrary individual
attacks, a secret key can be established in the SIMCAP scenario if
and only if the two-qubit state shared by Alice and Bob is
entangled. This gives a one-to-one correspondence between
two-qubit entanglement and secure key distribution: any one-qubit
channel that is not
entanglement breaking is secure. 
It would be interesting to extend our results to higher
dimensional systems (some preliminary results can be found in Ref.
\cite{AGS}), where there are entangled states, known as bound
entangled \cite{bound}, that are not quantum distillable. Our
analysis can be trivially extended to the so-called one-copy
distillable states, those states for which there exist local
projections onto two-dimensional subspaces such that the resulting
two-qubit state is entangled. The honest parties should simply
include these projections as a first step in the measurement part
of the distillation protocol. This fact suggests a complete
equivalence between distillable entanglement and key distribution.
According to it, the so-called entanglement binding channels,
those channels through which only bound entanglement can be
established \cite{binding}, would be useless for key distribution,
although this remains unproven. A related open question is the
conjectured existence of a classical analog of bound entanglement,
known as {\sl bound information} \cite{GW2}, that seems to appear
in some probability distributions $P(a,b,e)$ derived from bound
entangled states.

\medskip


We thank Dan Collins, Valerio Scarani and Stefan Wolf for
discussion. This work has been supported by the ESF, the Swiss
NCCR, ``Quantum Photonics" and OFES within the European project
RESQ (IST-2001-37559), the Spanish grant 2002FI-00373 UB and the
Generalitat de Catalunya.

\medskip

{\sl Appendix:} In this appendix we present several measurements
strategies for Eve that attain the exponential bound of Eq.
(\ref{expterm}). For all these strategies, Eqs. (\ref{eveerror}),
(\ref{eveerror2}) and (\ref{eveerror3}) become an equality in the
limit of large $N$. In other words, the r.h.s. of these equations
represent the relevant term of $\epsilon_{EN}$ when
$N\rightarrow\infty$. Thus, we only need to check Eq.
(\ref{expterm}) for the given measurements.

First, consider a projective measurement in the $x$ basis (see
Fig. \ref{figmeas}), i.e. $M_1=\proj{+x}$ and $M_2=\proj{-x}$.
This is the optimal measurement in terms of Eve's
information and error probability. 
Eve acquires information about $x$ from all the $N$ measurement
outcomes. Although we are not interested in her decision strategy,
she can associate $\ket{+x}$ ($\ket{-x}$) to $\ket{e_0}$
($\ket{e_1}$) and then apply a majority rule for guessing $x$. It
is now easy to see that
$\bra{e_0}M_1\ket{e_1}=\bra{e_0}M_2\ket{e_1}>0$, and therefore the
inequality (\ref{expterm}) is saturated by this measurement.

A second possibility corresponds to the measurement optimizing
Eve's probability of inconclusive result. It consists of three
operators, $M_1=c\,\proj{e_1^\perp}$, $M_2=c\,\proj{e_0^\perp}$
and $M_3=c_?\,\proj{+z}$. Note that if the first (second) outcome
is obtained, Eve knows that the state was $\ket{e_0}$
($\ket{e_1}$) with certainty, while she obtains no information
from $M_3$. The weights $c$ and $c_?$ are chosen such that the
probability of inconclusive result is minimized, giving
$p_?=\bra{e_0}M_3\ket{e_0}=\bra{e_1}M_3\ket{e_1}=|\langle
e_{1}\ket{e_{0}}|$. In this case it is simple to compute
$\epsilon_{EN}$ for all $N$. Knowing one of the symbols used in
$\vec a$ plus the information in $\vec x$ allows Eve to deduce
$x$. Only when she has obtained $N$ inconclusive results she is
forced to guess, making a mistake in half of the cases. Then, her
error probability reads $\epsilon_{EN}=\frac{1}{2}|\langle
e_{1}\ket{e_{0}}|^N$, which attains the bound.

Finally, all the measurements interpolating in a coherent (or
incoherent) way between these two strategies also attain the bound
(see Fig. \ref{figmeas}). Indeed it is simple to see that the
inequality (\ref{expterm}) is saturated by all of them. Let us
stress again here that this does not mean that $\epsilon_{EN}$ is
the same for all these measurements, but only that its exponential
behavior goes like $|\langle e_{1}\ket{e_{0}}|^N$ for large $N$,
i.e. $\lim_{N\rightarrow\infty}\,\log\epsilon_{EN}= N\log|\langle
e_{1}\ket{e_{0}}|$.


\begin{references}

\bibitem{book}
See for instance M. A. Nielsen and I. L. Chuang, {\sl Quantum
Computation and Quantum Information}, Cambridge University Press
(2000).






\bibitem{review}
N. Gisin {\sl et al.}, Rev. Mod. Phys {\bf 74}, 145 (2002).

\bibitem{BBM}
C. H. Bennett, G. Brassard and N. D. Mermin, Phys. Rev. Lett. {\bf
68}, 557 (1992).

\bibitem{Ekert}
A. Ekert, Phys. Rev. Lett. {\bf 67}, 661 (1991).

\bibitem{notesource}
Usually, Alice sends half of a maximally entangled state to Bob.
Here, we aim to discuss the most general
situation, with no constrains on $\rho_{AB}$. 
This is equivalent to the case where the state is prepared by an
insecure source.

\bibitem{noteind}
This assumption excludes unconditional security.

\bibitem{dist}
C. H. Bennett {\sl et al.}, Phys. Rev. Lett. {\bf 76}, 722 (1996);
D. Deutsch {\sl et al.}, Phys. Rev. Lett. {\bf 77}, 2818 (1996).

\bibitem{horo}
It was proven in M. Horodecki, P. Horodecki and R. Horodecki,
Phys. Rev. Lett. {\bf 78}, 574 (1997), that all two-qubit
entangled states are distillable.


\bibitem{optfilt}
A. Kent, N. Linden and S. Massar, Phys. Rev. Lett. {\bf 83}, 2656
(1999); F. Verstraete, J. Dehaene and B. DeMoor, Phys. Rev. A {\bf
64}, 010101(R) (2001).

\bibitem{bellbasis}
The Bell basis is defined by the four orthonormal two-qubit
maximally entangled states
$\ket{\Phi^{\pm}}=(\ket{00}\pm\ket{11})/\sqrt 2$ and
$\ket{\Psi^{\pm}}=(\ket{01}\pm\ket{10})/\sqrt 2$.


\bibitem{parttr}
A. Peres, Phys. Rev. Lett. {\bf 77}, 1413 (1996); M. Horodecki, P.
Horodecki and R. Horodecki, Phys. Lett. A {\bf 223}, 1 (1996).
Given an operator on $\compl^{d_1}\otimes\compl^{d_2}$, the
partial transposition of $O$ with respect to the first subsystem
in the basis $\{\ket{1},\ldots,\ket{d_1}\}$ is $O^{T_1}\equiv
\sum_{i,j=1}^{d_1}\bra{i}O\ket{j}\ket{j}\!\bra{i}$.

\bibitem{notemeas}
The filter plus the $z$ measurement can be seen as a single local
measurement of three outcomes: 0, 1 and reject.

\bibitem{GW}
N. Gisin and S. Wolf, Phys. Rev. Lett. {\bf 83}, 4200 (1999).

\bibitem{CK}
I. Csisz\'ar and J. K\"{o}rner, IEEE Trans. Inf. Theory {\bf
IT-24}, 339 (1978).

\bibitem{Maurer}
U.M. Maurer, IEEE Trans. Inf. Theory {\bf 39}, 733 (1993).

\bibitem{note2}
For large $N$, the first requirement is naturally satisfied by all
the typical sequences.


\bibitem{MW}
U. Maurer and S. Wolf, IEEE Trans. Inf. Theory {\bf 45}, 499
(1999).

\bibitem{GW2}
N. Gisin and S. Wolf, , {\em Proceedings of CRYPTO 2000}, Lecture
Notes in Computer Science {\bf 1880}, 482, Springer-Verlag, 2000,
quant-ph/0005042.

\bibitem{note}
It seems harder to do the same in the scheme using quantum
distillation, i.e. collective quantum operations.

\bibitem{HSR}
See M. Horodecki, P. W. Shor and M. B. Ruskai, quant-ph/0302031
and references therein.



\bibitem{AGS}
A. Ac\'\i n, N. Gisin and V. Scarani, quant-ph/0303009; D.
Bru{\ss}
 {\sl et al.}, quant-ph/0303184.

\bibitem{bound}
M. Horodecki, P. Horodecki and R. Horodecki, Phys. Rev. Lett. {\bf
80}, 5239 (1998).

\bibitem{binding}
P. Horodecki, M. Horodecki and R. Horodecki, J. Mod. Opt. {\bf
47}, 347 (2000).

\end{references}
\end{document}